\documentclass{article}

\usepackage{arxiv}

\usepackage[utf8]{inputenc} 
\usepackage[T1]{fontenc}    
\usepackage{microtype}      
\usepackage{authblk}        

\usepackage{hyperref}       
\usepackage{booktabs}       
\usepackage{multicol}       
\usepackage{makecell}       
\usepackage{amsfonts}       
\usepackage{amsmath}
\usepackage{graphicx}       
\usepackage{caption}
\usepackage[switch]{lineno} 

\newcommand{\SA}[0]{StrongAugment}
\newcommand{\RA}[0]{RandAugment}
\newcommand{\RAHE}[0]{$\text{RandAugment-HE}$}
\newcommand{\TA}[0]{TrivialAugment}

\title{Augment like there's no tomorrow: Consistently performing neural networks for medical imaging}

\author[1,$\dagger$,*]{Joona Pohjonen}
\author[1,$\dagger$]{Carolin Stürenberg}
\author[1]{Atte Föhr}
\author[3]{Reija Randen-Brady}
\author[1,3]{Lassi Luomala}
\author[3]{Jouni Lohi}
\author[2,4,5,$\ddagger$]{Esa Pitkänen}
\author[1,5,6,$\ddagger$]{Antti Rannikko}
\author[1,4,5,$\ddagger$]{Tuomas Mirtti}

\affil[1]{Research Program in Systems Oncology, Faculty of Medicine, University of Helsinki}
\affil[2]{Institute for Molecular Medicine Finland (FIMM), HiLIFE, University of Helsinki}
\affil[3]{Department of Pathology, Helsinki University Hospital}
\affil[4]{Research Program in Applied Tumor Genomics, Faculty of Medicine, University of Helsinki}
\affil[5]{iCAN Digital Precision Cancer Medicine Flagship, Finland}
\affil[6]{Department of Urology, Helsinki University Hospital}
\affil[$\dagger$]{Authors contributed equally}
\affil[$\ddagger$]{Senior author}

\begin{document}

\twocolumn[\begin{@twocolumnfalse}\maketitle\end{@twocolumnfalse}]

{
  \renewcommand{\thefootnote}{*}
  \footnotetext[1]{
        Correspondence: \texttt{joona.pohjonen@helsinki.fi} \\
        \hspace*{1.8em}Code: \url{https://github.com/jopo666/StrongAugment}
    }
}

\begin{abstract}
    \textbf{\normalsize Deep neural networks have achieved impressive performance in a wide variety of medical imaging tasks. However, these models often fail on data not used during training, such as data originating from a different medical centre. How to recognise models suffering from this fragility, and how to design robust models are the main obstacles to clinical adoption. Here, we present general methods to identify causes for model generalisation failures and how to circumvent them. First, we use \textit{distribution-shifted datasets} to show that models trained with current state-of-the-art methods are highly fragile to variability encountered in clinical practice and then develop a \textit{strong augmentation} strategy to address this fragility. Distribution-shifted datasets allow us to discover this fragility, which can otherwise remain undetected after validation against multiple external datasets. Strong augmentation allows us to train robust models achieving consistent performance under shifts from the training data distribution. Importantly, we demonstrate that strong augmentation yields biomedical imaging models which retain high performance when applied to real-world clinical data. Our results pave the way for the development and evaluation of reliable and robust neural networks in clinical practice.}
\end{abstract}


\section{Introduction}
\label{sec:introduction}

\begin{figure}
    \centering
    \includegraphics[width=\columnwidth]{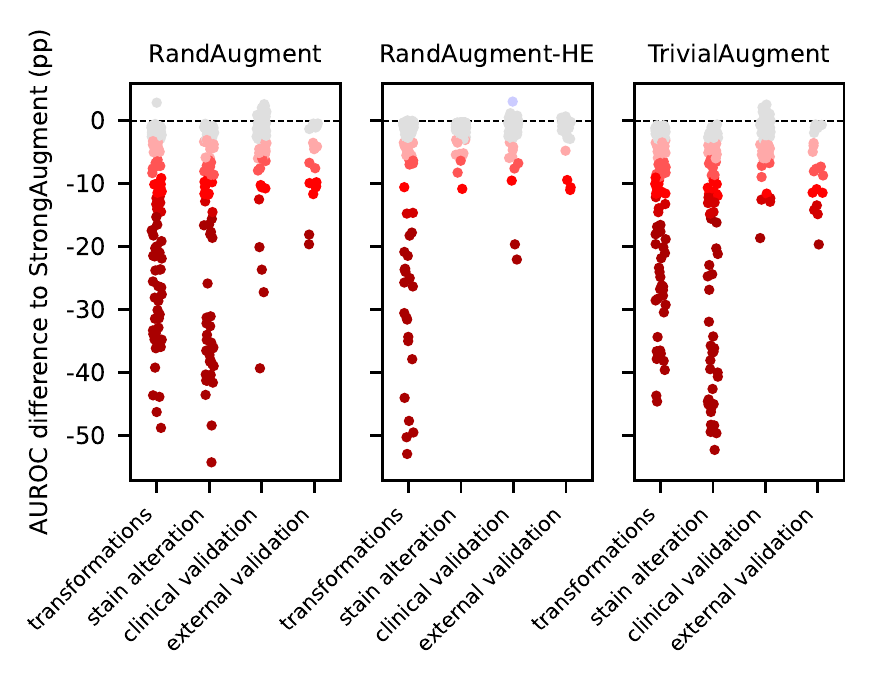}
    \caption{AUROC percentage-point (pp) difference between \SA{} and the current state-of-the-art methods on a cancer classification task. Neural networks trained with strong augmentation retain similar performance on all distribution-shifted datasets, external datasets and clinical validation cases, achieving up to 50 percentage points higher discrimination performance than current state-of-the-art training methods.}
    \label{fig:difference}
\end{figure}

Through digitalisation and neural network-based solutions, pathology is experiencing its third revolution \cite{third_revolution}. Neural networks have been applied to tissue diagnostics with impressive results, often surpassing human counterparts in consistency, speed and accuracy \cite{dl_success1, dl_success2}. Despite promising results, many recent works have demonstrated neural networks performing substantially worse on datasets not used during training \cite{campanella2019, nagpal2019, stain_norm_gan_3, liu2019_reduction, pneumonia, melanoma, covid_radiograph}. The two main challenges facing any neural network intended for clinical practice, stem from this generalisation problem \cite{to_clinic}. How to train neural networks which perform consistently despite real-world variability, and how to demonstrate this consistency?

All clinical-grade neural networks should be robust to possible data distribution shifts encountered in a clinical practice \cite{spectral_decoupling}. This robustness is hard to assess and has been commonly evaluated by performance on external validation datasets. Although an important part of neural networks' path to clinic \cite{to_clinic}, external validation is limited to exposing a failure to generalise in the particular dataset, and cannot establish clinical usefulness, robustness or generalisation ability of a network \cite{reporting_standard}. Even when a neural network achieves good performance on multiple external datasets, it may fail on another dataset or a small subset of samples. Thus, there is a critical need for evaluating neural networks' robustness to data distribution shifts. 

A promising solution for evaluating neural networks' robustness to distribution shifts is shifted evaluation \cite{shift_eval}. With shifted evaluation, a distribution-shifting function is applied to every sample in a dataset to create a new distribution-shifted dataset. The used function completely describes any performance differences between the original and distribution-shifted datasets and thus provides a measure of robustness as well as guidance on how to improve the evaluated model in case of failure. For example, distribution-shifted datasets created with the JPEG-compression algorithm \cite{jpeg} can be used to measure whether a neural network is robust to artefacts created during image compression. Shifted evaluation has been used in several recent works, to demonstrate that neural networks are extremely fragile to even small distribution shifts from the training data distribution \cite{spectral_decoupling, stress_testing, shifted_evaluation_1, shifted_evaluation_2}.

Large multi-cohort datasets, which are representative of all types of data encountered in clinical practice, are often seen as the main solution to the generalisation problem \cite{to_clinic}. Although important, larger datasets often have lower data quality \cite{dl_failures} and different cohorts introduce their own biases, which makes training harder as neural networks overfit these biases easily \cite{covid_radiograph, spectral_decoupling, shortcut_learning}. Still, even a large high-quality multi-cohort dataset could not fully represent all types of data encountered in clinical practice, and other complementary methods for improving neural networks' robustness are needed. Another popular method is stain normalisation, which reduces the variability (mainly colour) by normalising each image to a well-defined common standard \cite{stain_norm_gan_3, macenko, stain_norm_autoencoder, stain_norm_gan, stain_norm_gan_2, stain_norm_and_augment, stain_norm}. Although stain normalisation can be useful, it is often computationally costly and does not fix the underlying problem of neural networks' fragility towards distribution shifts \cite{stress_testing}.

Only a few solutions exist for increasing the robustness to distribution shifts from the training data. Recently proposed method of spectral decoupling \cite{gradient_starvation} can be used to increase robustness to data distribution shifts, as well as avoid overfitting unwanted biases, for example in multi-cohort datasets \cite{spectral_decoupling}. Another solution is to introduce artificial variability to the images during neural network training, also known as augmentation. Augmentation has been shown to explicitly improve robustness to certain data distribution shifts \cite{spectral_decoupling, stain_norm_and_augment, stain_augment}. For example, the sharpness of the scanned pathology slide images varies significantly between medical centres, scanning equipment and slide areas \cite{spectral_decoupling, stain_norm_and_augment}. By randomly adjusting the sharpness of the images during training, neural networks can be made robust to sharpness variations. Automatic augmentation strategies \cite{autoaugment, randaug, trivaug} have become a standard in training neural networks \cite{resnetrs}. However, augmentation has been traditionally seen as a way to increase the amount of training data, and the development of augmentation strategies has not focused explicitly on improving neural networks' robustness to data distribution shifts.

In this study, we propose general methods for evaluating and training robust neural networks intended for clinical practice. First, we demonstrate how shifted evaluation can be used to evaluate neural networks' robustness to data distribution shifts, and thus also its generalisation ability in clinical practice. We discover that neural networks trained with the current state-of-the-art methods are extremely fragile to even small shifts from the training data distribution. We then develop an automatic augmentation strategy, \SA{}, to address this fragility. Using large-scale heterogeneous histopathology data from several studies, we demonstrate that strong augmentation yields robust biomedical imaging models, while state-of-the-art methods achieve significantly lower discrimination performance (Figure \ref{fig:difference}).

\section{Materials and methods}
\label{sec:materials_methods}

\subsection{Shifted evaluation with distribution-shifted datasets}
\label{sec:shifted_evaluation}

Let $x_1, \dots, x_n \in X$ denote a dataset with an unknown data generating process $\mathcal{P}$. To create a distribution shifted dataset $f(X)$, the distribution shifting function $f$ is applied to each image $x_i \in X$, where $i \in 1, \dots, n$. Although the data generating process $\mathcal{P}$ is unknown, any performance differences between the datasets $f(X)$ and $X$ are guaranteed to be due to the distribution shift defined by $f$. For example, if a given model achieves an accuracy of 0.9 for dataset $X$ and 0.5 for dataset $f_{invert}(X)$, where $f_{invert}(x_i) = 255 - x_i$, we can say that the given model is sensitive to the distribution shift caused by colour inversion.

The data generating process $\mathcal{P}$ is inherently unknown for external datasets, as the image acquisition process cannot be exhaustively described. Thus, even when a given model passes external validation, it cannot establish clinical usefulness, robustness or generalisation ability of the model as there is no way of knowing why the model succeeds. External validation can only expose a failure to generalise, with no feedback given to the researcher on reasons for the failure.

\subsubsection{Image transformations}
\label{sec:data_trnsf}

\begin{figure*}
    \centering
    \includegraphics{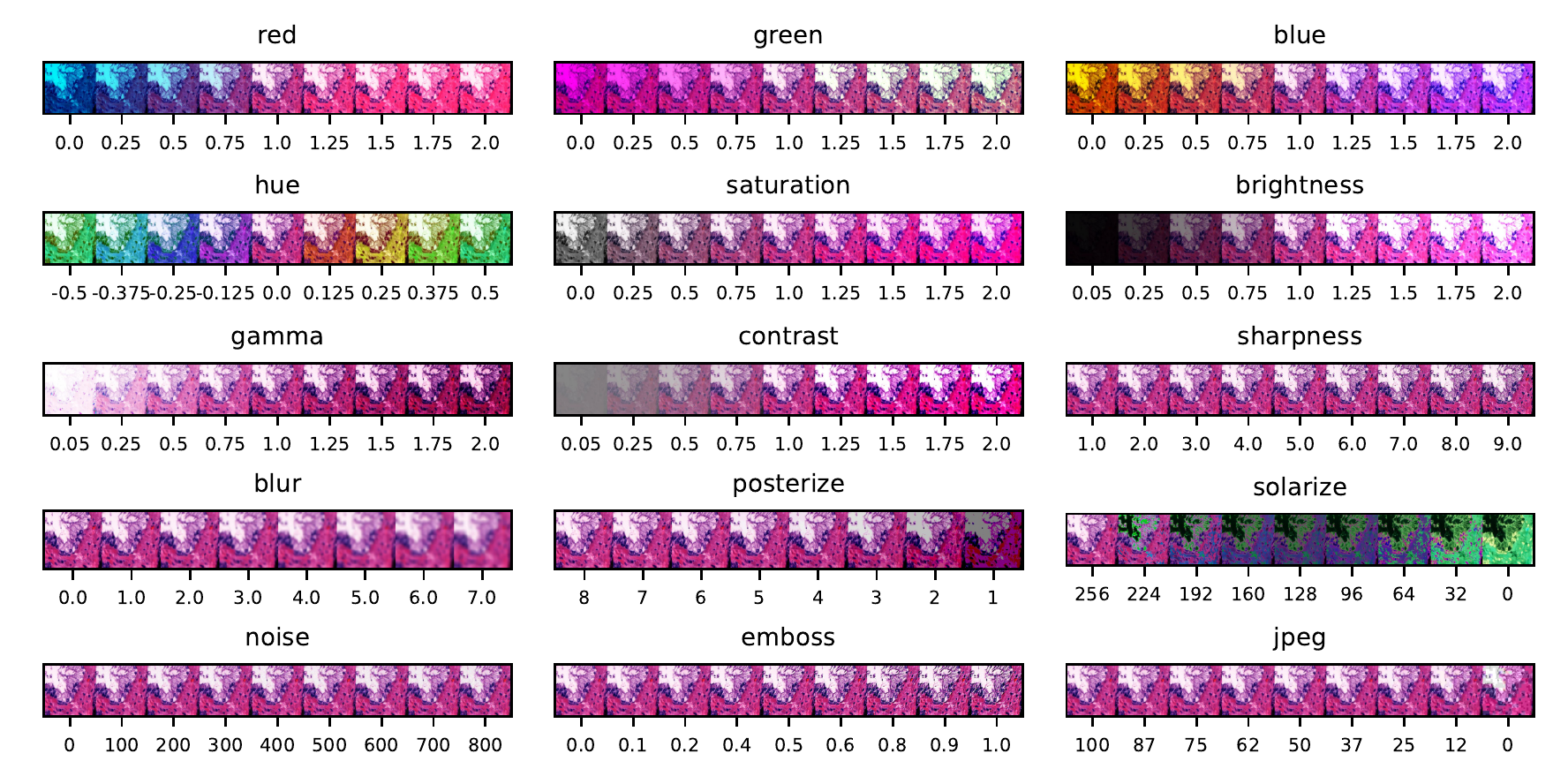}
    \caption{Examples of image transformations with $t(x_i, m)$, where transformation $t$ is denoted by the figure title and magnitude $m$ by the x-axis value.}
    \label{fig:transformations}
\end{figure*}

To evaluate neural networks' overall robustness to data distribution shifts, 15 image transformations are used to create distribution-shifted datasets. These datasets are denoted by $t(X, m)$, where $t$ denotes the used transformation and $m$ its magnitude. By increasing or decreasing magnitude $m$, the data distribution can be incrementally shifted further from the original distribution of dataset $X$. Figure \ref{fig:transformations} presents all selected transformations with examples of transformed images with different magnitudes.

The transformations and magnitudes, presented in Figure \ref{fig:transformations}, were selected to create distribution-shifted datasets to evaluate neural networks' overall robustness to data distribution shifts. It should be reasonable to expect a neural network to perform similarly on all of the images in Figure \ref{fig:transformations}, and any potential failures will give important insights into the robustness of the model. Although some of the distribution-shifted datasets may not be directly encountered in routine clinical practice, they can still be useful in other ways. For example, if a given network performs consistently for dataset $\textit{posterize}(X, 4)$, $\textit{saturation}(X, 0)$ and $\textit{jpeg}(X, 10)$, it could be possible to significantly reduce the size of large digital slides by reducing the number of bits, dropping colour channels or compressing images more heavily, without compromising model performance.

\textit{Noise}, \textit{emboss} and \textit{jpeg} transformations are from the \texttt{albumentations} (1.2.1) library \cite{albumentations}, \textit{red}, \textit{green}, \textit{blue} transformations have been defined by us, and all other transformations are from the \texttt{torchvision} (1.11.1) library \cite{torchvision}.

\subsubsection{Haematoxylin and eosin stain intensities}
\label{sec:data_stain}

Colours of the digital slide images can vary substantially due to differences in the staining processes and the choice of imaging equipment and its settings, especially between medical centres \cite{stain_norm_and_augment}. Although several transformations presented in Figure \ref{fig:transformations} evaluate robustness to changes in the colour of the images, none of them explicitly measure robustness to data distribution shifts caused by differences in the staining process. For this reason, and to demonstrate the flexibility of distribution-shifted datasets, distribution-shifted datasets are created by adjusting the intensity of the haematoxylin and eosin stains.

\begin{figure}
    \centering
    \includegraphics[width=\columnwidth]{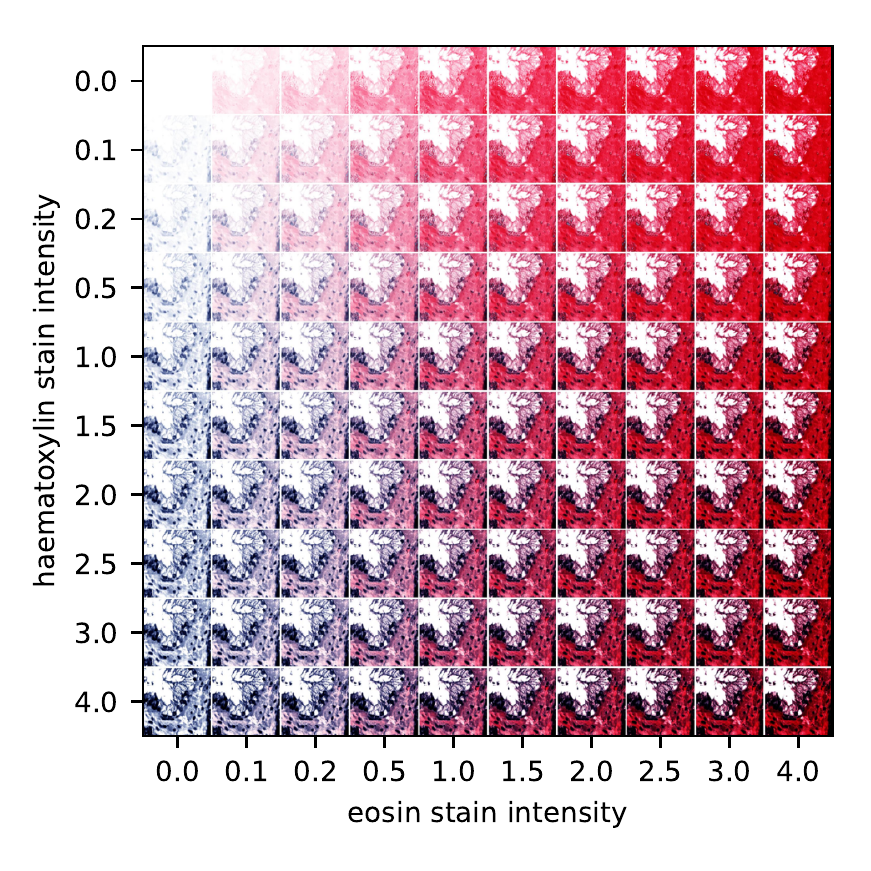}
    \caption{Examples of haematoxylin and eosin stain intensity adjustments where haematoxylin and eosin staining intensities are denoted by the y-axis and x-axis, respectively.}
    \label{fig:stain_example}
\end{figure}

To create a distribution-shifted dataset $\textit{stain}(X, h, e)$, the Macenko method \cite{macenko} is used to separate the haematoxylin and eosin stains, and then concentrations of each stain are multiplied with magnitudes $h$ and $e$, respectively. For example, $h=0.0, e=1.0$ would remove the eosin stain completely from the image and $e=1.0, m_h=2.0$ would double the haematoxylin stain intensity. Figure \ref{fig:stain_example} demonstrates example images from the distribution-shifted datasets. The mean stain vector of the original dataset is used to separate the stains.

\subsection{StrongAugment}
\label{sec:strong_augment}

\begin{table*}[t]
    \centering
    \caption{Augmentation spaces of \SA{}, \RA{} \cite{randaug}, \RAHE{} \cite{randaug_he} and \TA{} \cite{trivaug}. No effect column represents the magnitude which does not change the image where applicable. \SA{} includes 6 additional transformations, Gaussian blurring and wider magnitude ranges for most of the common transformations. Examples for some of the transformations can be seen in Figure \ref{fig:transformations}. Please note, that a hyper-parameter is used to further limit the magnitude ranges for \RA{} and \RAHE{}.}
    \begin{tabular}{llllll}
Transformation   & \SA{} & \RA{} & \RAHE{} & \TA & No effect \\
\midrule
Identity         & included         & included        & included         & included         &    \\
Shear\{X,Y\}     & {[}-145, 145{]}  & {[}-17, 17{]}   & {[}-51, 51{]}    & {[}-145, 145{]}  & 1.0\\
Translate\{X,Y\} & {[}-32, 32{]}    & {[}-72, 72{]}   & {[}-30, 30{]}    & {[}-32, 32{]}    & 0  \\
Rotate           & {[}-135, 135{]}  & {[}-30, 30{]}   & {[}-90, 90{]}    & {[}-135, 135{]}  & 0  \\
Saturation       & {[}0.0, 2.0{]}   & {[}0.1, 1.9{]}  & {[}0.0,  5.5{]}  & {[}0.01, 1.99{]} & 1.0\\
Brightness       & {[}0.1, 1.9{]}   & {[}0.1, 1.9{]}  & {[}0.0,  5.5{]}  & {[}0.01, 1.99{]} & 1.0\\
Contrast         & {[}0.1, 1.9{]}   & {[}0.1, 1.9{]}  & {[}0.0,  5.5{]}  & {[}0.01, 1.99{]} & 1.0\\
Sharpness        & {[}1.0, 2.0{]}   & {[}0.1, 1.9{]}  & {[}0.0,  5.5{]}  & {[}0.01, 1.99{]} & 1.0\\
Gaussian blur    & {[}0.0, 2.0{]}   & –               & –                & –                & 0.0\\
Solarize         & {[}0, 255{]}     & {[}0, 255{]}    & –                & {[}0, 255{]}     & 256\\
Posterize        & {[}1, 8{]}       & {[}4, 8{]}      & –                & {[}1, 8{]}       & 8  \\
Equalize         & included         & included        & included         & included         &    \\
Auto contrast    & included         & included        & included         & included         &    \\
\midrule
Grayscale        & included         & –               & –                & –                &    \\
Gamma            & {[}0.1, 1.9{]}   & –               & –                & –                & 1.0\\
Hue              & {[}-0.5, 0.5{]}  & –               & –                & –                & 0.0\\
Red              & {[}0.01, 1.99{]} & –               & –                & –                & 1.0\\
Green            & {[}0.01, 1.99{]} & –               & –                & –                & 1.0\\
Blue             & {[}0.01, 1.99{]} & –               & –                & –                & 1.0\\
\midrule
HED colour shift & –                & –               & {[}-0.9, 0.9{]}  & –                & 0.0\\
HSV colour shift & –                & –               & {[}-0.9, 0.9{]}  & –                & 0.0\\
\bottomrule
\multicolumn{5}{l}{\footnotesize{HED: hematoxylin-eosin-DAB, HSV:  and hue-saturation-value}}
    \end{tabular}
    \label{tab:augment_space}
\end{table*}

In this study, we create an automatic augmentation method, \SA{}, which is designed to increase neural networks' robustness to data distribution shifts from the training data. \SA{} is compared to \RA{} \cite{randaug}, one of the most commonly used automatic augmentation methods, and \TA{} \cite{trivaug} which represents the current state-of-the-art in automatic augmentation. Importantly, implementations for both methods are easily accessible through the \texttt{torchvision} library, and thus likely used by other researchers in the field. Additionally, we compare \SA{} to a HE-tailored variation of \RA{}, \RAHE{} \cite{randaug_he}, which exploits prior knowledge of the presumable variation of histological images such as differences in the haematoxylin and eosin stains seen in Figure \ref{fig:stain_example}. This allows comparisons between augmentation strategies explicitly designed to increase robustness for certain data distribution shifts, such as HE-staining differences, and \SA{}, which is designed to increase robustness for all distribution shifts.

\SA{} applies a varying number of transformations sequentially to an image $x$. After $x$ has been transformed two times, each consecutive transformation is applied with probability $p$, or until $x$ has been transformed five times. Each transformation $t$ and its magnitude $m$ is sampled uniformly from an augmentation space $A$, which contains a specified list of transformations and a range of possible magnitudes for each transformation. After applying transformation $t$, the transformation $t$ is removed from $A$. Only one affine transformation is allowed per image, to avoid cases where sequentially applied affine transformations move the image data out of the image. This allows significantly higher number of transformations than previous state-of-the-art augmentation methods \cite{randaug, trivaug, randaug_he, uniformaug}.

Table \ref{tab:augment_space} represents the augmentation spaces of \SA{}, \RA{}, \RAHE{} and \TA{}, and Supplementary Figure \ref{fig:augment_comparison} examples of augmented images with each method. \SA{} has both a wider augmentation space $A$ and a higher number of transformations applied per image. The augmentation space of \SA{} contains all the same transformations as \RA{} and \TA{}, with six additional transformations. To enable stronger blurring of images, the default degenerate method implemented by the \texttt{torchvision} library was replaced with Gaussian blurring. No neural networks were trained while building the augmentation space to avoid overfitting the augmentation space to a specific dataset. Transformations and their magnitude ranges were simply chosen to be as wide as possible.

\RA{}, \TA{} have been implemented by \texttt{torchvision} (1.11.1). \RAHE{} has been implemented in a previous study \cite{randaug_he}. For \RA{}, the magnitude $m$ and number of operations $n$ are set to $m=10$ and $n=2$ \cite{resnetrs}, and for \RAHE{} to $m=5$ and $n=3$ \cite{randaug_he}. For \TA{} there are no tunable hyper-parameters.

\subsection{Training details}
\label{sec:training_details}

ResNet-RS50 \cite{resnetrs} neural networks are used for all the experiments in this study. Dropout \cite{dropout} and stochastic depth \cite{stochastic_depth} are set to 0.2 \cite{resnetrs} for all networks. For training, a random area of the image is cropped and resized to $160 \times 160$ pixels with either bi-linear, bi-cubic or nearest interpolation, and flipped horizontally and/or vertically with a probability of 0.5. Then the image is transformed with either \SA{}, \RA{}, \RAHE{} or \TA{}, a random area of the image is erased with an 0.2 probability, and either Mixup \cite{mixup} or Cutmix \cite{cutmix} is applied for pair of images in a batch with an alpha value of 0.2. For testing, the images are resized to $224 \times 224$ pixels with bi-linear interpolation. 

Adaptive sharpness aware minimisation \cite{asam} is used with $\rho=0.05$ and Adam \cite{adam} with $\beta_1 = 0.9$, $\beta_2 = 0.999$ as the base optimiser. The learning rate is increased linearly from $10^{-6}$ to $0.0004 \times \texttt{batch\_size}/256$ during the first 10 epochs and then reduced back to $10^{-6}$ with a cosine schedule. Label smoothing is set to 0.1 \cite{label_smoothing}. All networks are trained for 150 epochs with a batch size of 1024. For each experiment, five neural networks are trained and the mean and standard deviation of the evaluation metric are reported.

Spectral decoupling \cite{gradient_starvation} is used instead of weight decay, which has been shown to increase neural networks' robustness for data distribution shifts \cite{spectral_decoupling}. Spectral decoupling coefficient is set to 0.0001 and weight decay to 0. For experiments, where spectral decoupling is intentionally left out to assess its effect on neural networks' robustness, weight decay is set to 0.00004 \cite{resnetrs}.

PyTorch (1.10.0) (\cite{torch}) is used for training the neural networks, and ResNet-RS50, Mixup and Cutmix implementations are from the \texttt{timm} (0.5.4) library \cite{timm}.

\subsection{Evaluation metric}
\label{sec:eval_metric}

To evaluate the performance of a neural network in a given dataset, the area under the receiver operating characteristic curve (AUROC) is reported for each experiment. AUROC measures the ability to discriminate between positive and negative samples. The AUROC value corresponds to the proportion of positive and negative sample pairs, where the positive sample has a higher probability given by the classifier. AUROC values of 0.5 and 1.0 correspond to random and perfect discrimination, respectively.

\subsection{Datasets}
\label{sec:datasets}


\begin{table*}[t]
    \centering
    \caption{Datasets used in this study.}
    \begin{tabular}{llllll}
    \toprule
        Name                & Medical centre                    & Years     & Images    & Label noise & Reference \\
        \midrule
        Helsinki30          & Helsinki University Hospital      & 2014-2015 & 4.7 mil.  & 0.03\%       & \cite{spectral_decoupling} \\
        Helsinki60          & Helsinki University Hospital      & 2019-2020 & 13.1 mil. & 17.0\%      & – \\
        Karolinska          & Karolinska Institutet             & 2012-2014 & 0.6 mil.  & 9.6\%       & \cite{panda} \\
        Radboud             & Radboud University Medical Center & 2012-2017 & 0.8 mil.  & 3.9\%       & \cite{panda} \\
        PESO                & Radboud University Medical Center & 2006-2011 & 5655      & –           & \cite{peso_dataset, peso_study} \\
        Gleason2019         & Vancouver General Hospital        & 1997-2011 & 94230     & 14.7\%      & \cite{gleason2019} \\
    \midrule
        HelsinkiRCC      & Helsinki University Hospital      & 2006-2013 & 2.0 mil.  & 8.1\%       & – \\
    \bottomrule
    \end{tabular}
    \label{tab:datasets}
\end{table*}

Seven different datasets from two tissue types, originating from four countries are used in this study. A basic summary of the datasets is presented in Table \ref{tab:datasets}. All digital slide images are processed with the \texttt{HistoPrep} library (1.0.7) \cite{histoprep}.

A total of 30 prostate cancer patients' full glass slide sets from surgical specimens are annotated for classification into cancerous and benign tissue, where the cancerous areas were annotated in consensus by two observers (C.S., T.M.). All patients have undergone radical prostatectomy at the Helsinki University Hospital between 2014 and 2015. Each case contains 14 to 21 tissue section slides. Tissue sections have a thickness of 4 $\mu$m and were stained with haematoxylin and eosin in a clinical-grade laboratory at the Helsinki University Hospital Diagnostic Center, Department of Pathology. Two different scanners are used to obtain images of the tissue section slides at 20x magnification. Larger macro slides (whole-mount, 2x3 inch slides) are scanned with an Axio Scan Z.1 scanner (Zeiss, Oberkochen, Germany), and the normal size slides with a Pannoramic Flash III 250 scanner (3DHistech, Budapest, Hungary). From the 30 patient cases, seven are set aside for a validation set. Digital slide images are cut into tiles with $1024 \times 1024$ pixels and 20\% overlap, resulting in 4.7 million tiles with 10\% containing cancerous tissue. For training, we selected all cancerous tiles and sample randomly the same amount of benign tiles to have equal amounts of positive and negative samples. We denote this dataset as Helsinki30.

Another set of 60 radical prostatectomy slides is also annotated into cancerous and benign tissue by one of six experienced pathologists as part of routine clinical diagnostics. All patients have undergone radical prostatectomy at the Helsinki University Hospital between 2019 and 2020. Each case contains 10 to 21 normal and macro tissue section slides of the prostate. Tissue sections have a thickness of 4 $\mu$m and are also stained with haematoxylin and eosin in a clinical-grade laboratory at the Helsinki University Hospital Diagnostic Center, Department of Pathology. All slides are scanned with an Axio Scan Z.1 scanner. From the 60 patient cases, seven are set aside for a validation set. Digital slide images are cut into $1024 \times 1024$ pixel tiles with 20\% overlap, resulting in 13.1 million tiles with 16\% containing cancerous tissue. For training, we selected all cancerous tiles and sample randomly the same amount of benign tiles to have equal amounts of positive and negative samples. We denote this dataset as Helsinki60.

As external datasets, four publicly available prostate cancer datasets are used. These datasets yield five external datasets from three different countries.

The PANDA development dataset contains 10616 prostate biopsy slides from 2113 patients from Radboud University Medical Center between 2012 and 2017 and Karolinska Institutet (Stockholm, Sweden) between 2012 and 2014 \cite{panda}. The digital slide images are cut into $512 \times 512$ pixel tiles with 20\% overlap resulting in 609180 (48\% cancer) and 780694 (23\% cancer) tiles from Radboud University Medical Center and Karolinska Institutet, respectively. Tile images were labelled as cancerous if there was any overlap with the cancerous annotations. Validation splits are omitted for both datasets as patient identifiers are missing from the publicly available data. We denote the datasets from Radboud University Medical Center and Karolinska Institutet as Radboud and Karolinska, respectively.

The PESO dataset contains tissue section slides from patients who have undergone a radical prostatectomy at the Radboud University Medical Center (Nijmegen, the Netherlands) between 2006 and 2011 (\cite{peso_dataset, peso_study}). The dataset contains images with $2500 \times 2500$ pixels annotated by a uropathologist as either cancerous or benign. These images are cut into $512 \times 512$ pixel tiles with 20\% overlap, resulting in 5655 tiles with 45\% containing cancerous tissue.

The Gleason2019 dataset contains 333 prostate tissue microarray spots from 231 patients who had undergone radical prostatectomy at Vancouver General Hospital between 1997 and 2011 \cite{gleason2019}. These digital slide images are cut into $224 \times 224$ pixel tiles with 20\% overlap resulting in 94230 tiles with 88.5\% containing cancerous tissue.

To evaluate augmentation strategies on training data from a different tissue type, a dataset was compiled from 167 patients with renal cell carcinoma (clear cell). All patients had undergone radical nephrectomy at the Helsinki University Hospital between 2006 and 2013. Tissue sections were stained with haematoxylin and eosin in a clinical-grade laboratory at the Helsinki University Hospital Diagnostic Center, Department of Pathology. A total of 698 cancerous and 172 benign tissue microarray spots were gathered from the sections and scanned with Pannoramic Flash III 250 scanner (3DHistech, Budapest, Hungary). Of the 167 patients, 24 were set aside for a validation set. Scanned tissue microarray spots were cut into tiles with $384 \times 384$ pixels and 20\% overlap, resulting in 2.0 million tiles with 21\% containing cancerous tissue. We denote this dataset as HelsinkiRCC.

\subsubsection{Label noise}

Each dataset used in this study has a variable annotation strategy as they stem from different projects. Thus, each dataset contains different amounts of label noise. For this reason, the reported performances on different datasets are not directly comparable. For example, a perfect classifier would achieve 100\% accuracy on a clean dataset, but only 80\% accuracy on a dataset with 20\% label noise. In reality, there are no perfect classifiers \cite{no_free_lunch} and the decreased performance on an external dataset may also be caused by the fact that the classifier cannot generalise to the dataset. For this reason, the amount of label noise in a dataset should be estimated.

To estimate the amount of label noise in a given dataset, a neural network is trained on the dataset, and the training discrimination performance is reported. Due to overfitting, the estimate of label noise in the dataset can easily be deflated. To minimise overfitting, we use strong regularisation outlined in Section \ref{sec:training_details} with \SA{}. To evaluate the accuracy of this estimation method, label noise is introduced to the Helsinki30 dataset, which contains the least amount of label noise based on a niche expert evaluation (T.M.). After flipping 0, 5, 10 and 20\% of the labels in the Helsinki30 dataset, the estimated label noise is 0.03, 5.5, 10.21 and 20.16\%, respectively.

The estimated amount of label noise in the Helsinki30, Helsinki60, Karolinska, Radboud, Gleason2019 and HelsinkiRCC datasets are 0.03\%, 17.0\%, 9.6\%, 3.9\%, 14.7\% and 8.1\%, respectively.

\subsection{Clinical Validation}
\label{sec:clinical_val}

To evaluate neural networks' robustness to distribution shifts encountered in routine clinical practice, several clinical validation cases were collected. Three surgical specimen paraffin blocks from patients who have undergone radical prostatectomy at the Helsinki University Hospital (HUS) in June of 2022, were collected. For each specimen, 36 serial tissue sections were stained in one of four HUS Diagnostic Centre's clinical-grade laboratories, using one of three slide types. Each slide is then scanned with two different scanners at 20x magnification. The four medical centres are Hyvinkää Hospital (Hyvinkää, Finland), Jorvi Hospital (Espoo, Finland), Kotka hospital (Kotka, Finland) and Meilahti Hospital (Helsinki, Finland). The three microscope slides are Superfrost Plus (SP) (Fisher Scientific, Waltham, MA, United States), TOMO (Matsunami Glass, Bellingham, WA, United States) and Klinipath (KP) (VWR International, Radnor, PA, United States). The two scanners are Pannoramic 1000 (3DHistech, Budapest, Hungary) and Aperio AT2 (Leica Microsystems, Wetzlar, Germany). This produces 72 scanned slide images, which are annotated for classification into cancerous and benign tissue, by one observer (T.M.). There is considerable variability between medical laboratories, slide glasses and scanners, which is demonstrated in Supplementary Figure \ref{fig:clinval_slides}. 

\section{Results}
\label{sec:results}

\begin{figure*}
    \centering
    \includegraphics{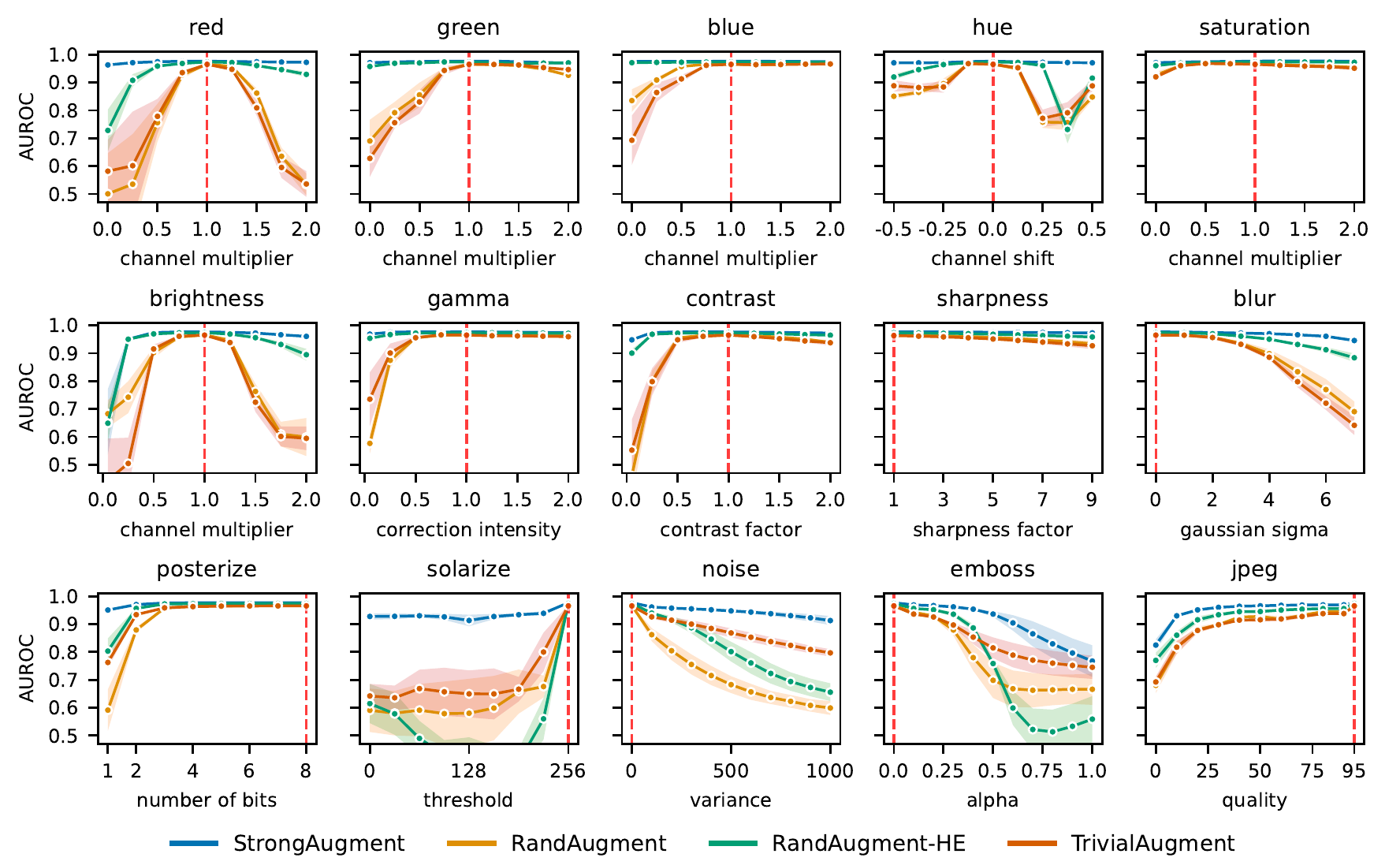}
    \caption{Robustness of neural networks, trained on the Helsinki30 dataset with either \SA{}, \RA{}, \RAHE{} or \TA{}, to distribution shifts caused by image transformations. The lines show the mean AUROC values and shaded regions with one standard deviation around the mean for the five trained networks. The distribution-shifted datasets $t(X_{\text{PESO}},m)$ are represented by dots, where the title denotes the transformation $t$ and the x-axis the magnitude $m$. Networks trained with \SA{} retain their discrimination performance despite increasingly large distribution shifts, whereas the performance of networks trained with \RA{} or \TA{} degrade quickly with even small distribution shifts. This fragility is not discoverable by simply looking at the performance on the unmodified PESO dataset, denoted by dotted red lines.}
    \label{fig:robustness_trnsf}
\end{figure*}
\begin{figure*}
    \centering
    \includegraphics{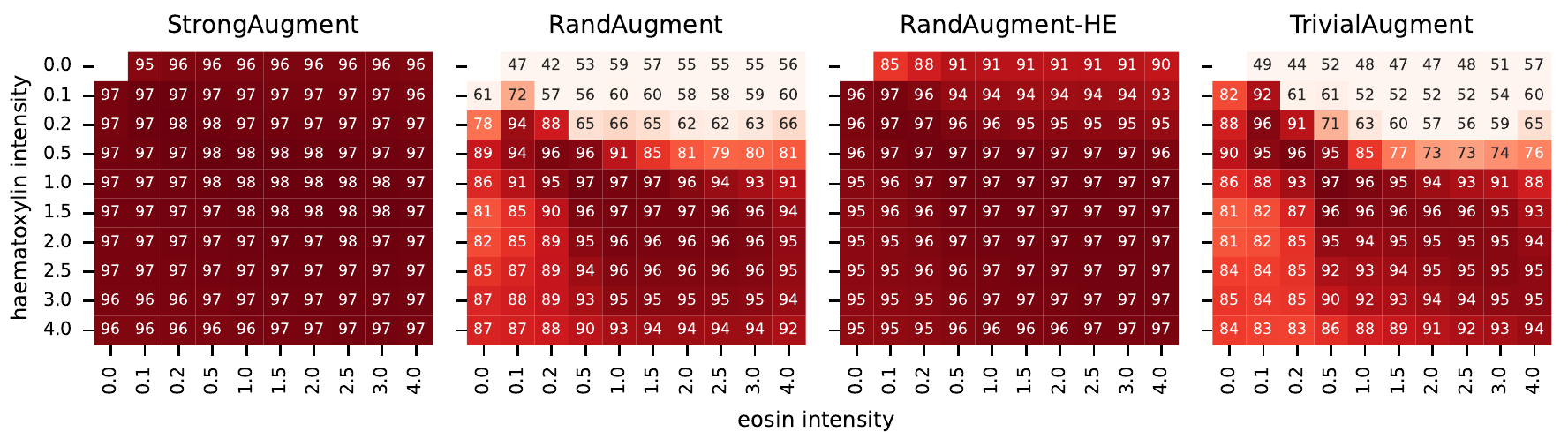}
    \caption{Robustness of neural networks, trained on the Helsinki30 dataset with either \SA{}, \RA{}, \RAHE{} or \TA{}, to distribution shifts caused by differences in the haematoxylin and eosin stain intensities. Each cell represents a given distribution-shifted dataset $\textit{stain}(X_{\text{PESO}}, h, e)$, where $h$ and $e$ denote the magnitudes for haematoxylin and eosin stains, respectively. Cells are annotated with mean AUROC $ \times$ 100 value across the five trained networks, rounded to the closest integer. Neural networks trained with \SA{} retain performance on all datasets, even when networks trained with \RA{} or \TA{} have degraded to no better than random discrimination. \SA{} also outperforms \RAHE{}, even though the latter method explicitly controls for staining differences.}
    \label{fig:robustness_stain}
\end{figure*}

\subsection{Robustness to data distribution shifts}
\label{sec:res_robustness}

First, neural networks' robustness to data distribution shifts is evaluated with distribution-shifted datasets created from the PESO dataset. For each experiment, five neural networks are trained on the Helsinki30 dataset with either \SA{}, \RA{}, \RAHE{} or \TA{} while keeping all other hyper-parameters fixed.

First, the general robustness to data distribution shifts is evaluated using the distribution-shifted datasets, described in Section \ref{sec:data_trnsf}, with examples presented in Figure \ref{fig:transformations}. Discrimination performances of the trained neural networks on these datasets are presented in Figure \ref{fig:robustness_trnsf}. Although all methods achieve similar performance on the original dataset, neural networks trained with either \RA{}, \RAHE{} or \TA{} are highly sensitive to even small data distribution shifts, which is evident from the large performance differences between the original and distribution-shifted datasets. Neural networks trained with \SA{} retain similar or significantly higher discrimination performances for all distribution-shifted datasets, even in cases where networks trained with other methods have degraded to no better than random discrimination. Notably, neural networks trained with \SA{} are robust to distribution shifts not included in the augmentation space, such as Noise, Emboss and JPEG transformations. Comparably, neural networks trained with either \RA{}, \RAHE{} or \TA{} are sensitive to many of the transformations included in their augmentation space, such as Hue and Brightness transformations.

Second, neural networks' robustness to data distribution shifts caused by differences in the haematoxylin and eosin stain intensities is evaluated using the distribution-shifted datasets described in Section \ref{sec:data_stain}, with example images presented in Figure \ref{fig:stain_example}. Discrimination performances of the trained neural networks on these datasets are presented in Figure \ref{fig:robustness_stain}. Neural networks trained with either \RA{}, or \TA{} are highly sensitive to distribution shifts caused by changes in the haematoxylin and eosin stain intensities, and the discrimination performance degrades to no better than random with even small distribution shifts. Neural networks trained with \SA{} retain their discrimination performance on all distribution-shifted datasets, even with datasets where networks trained with \RA{} or \TA{} have no better than random discrimination. \SA{} also significantly outperforms \RAHE{}, which explicitly controls for the distribution shifts caused by haematoxylin and eosin differences \cite{randaug_he}.

We recommend comparing the discrimination performances in Figures \ref{fig:robustness_trnsf} and \ref{fig:robustness_stain} to the example images in Figures \ref{fig:transformations}, \ref{fig:stain_example}, respectively.

\subsection{External validation}
\begin{figure*}[ht]
    \centering
    \includegraphics{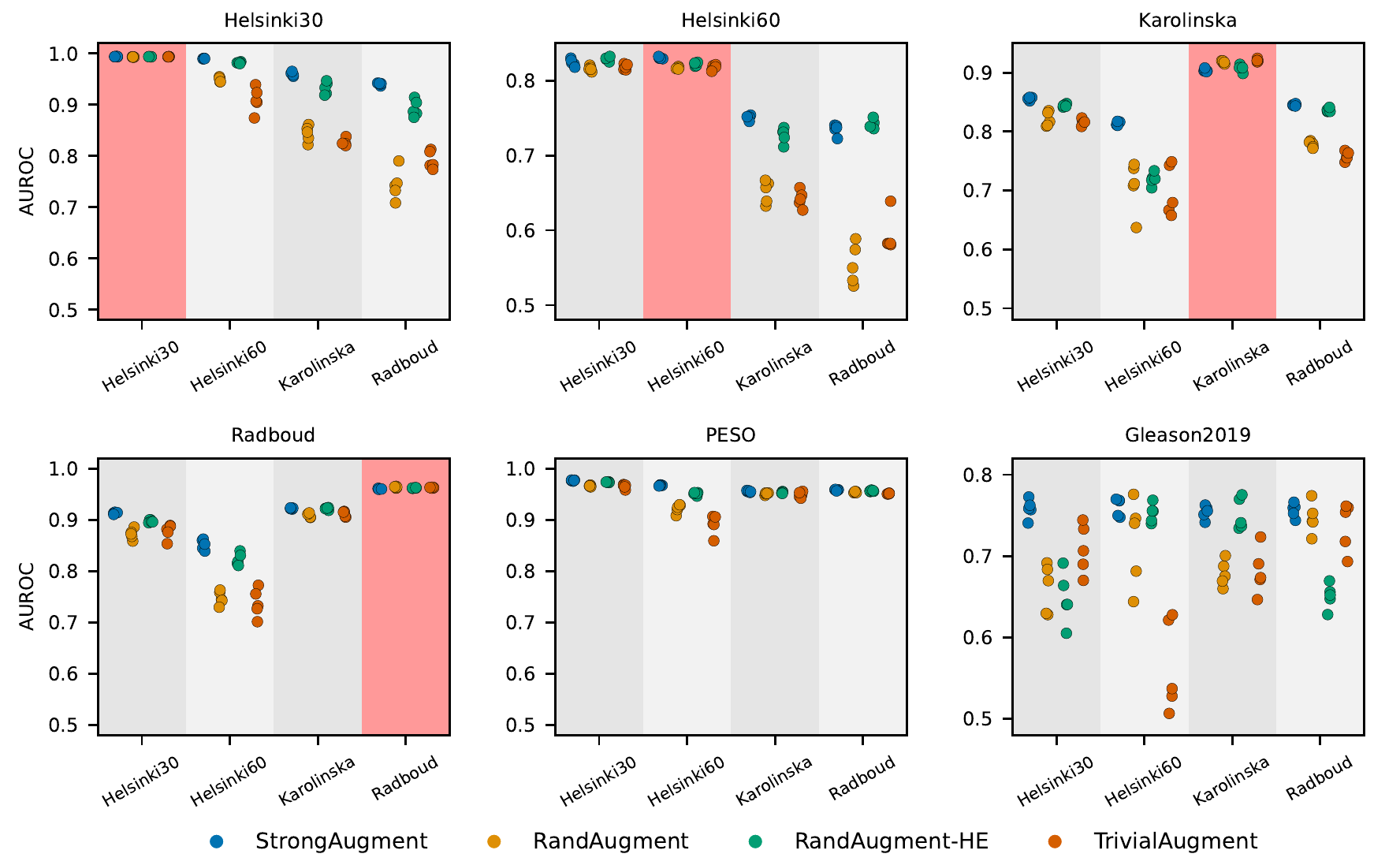}
    \caption{Discrimination performances of neural networks trained with \SA{}, \RA{}, \RAHE{} or \TA{} on four different training datasets. Each network is evaluated with five external datasets and the training dataset, denoted by a red box. Evaluation datasets are denoted by the figure titles and training datasets by the x-axis labels. Neural networks trained with \SA{} achieve better or comparable performance on every evaluation dataset, whereas \RA{}, \RAHE{} and \TA{} demonstrate a significant lack of generalisation ability with several evaluation datasets. Please note that the y-axes of Helsinki60, Karolinska and Gleason2019 evaluation datasets have been truncated to account for the label noise in these datasets.}
    \label{fig:external_val}
\end{figure*}

To assess, whether better performance on the distribution-shifted datasets translates to better performance on real-world datasets, we evaluate the trained neural networks on the Helsinki30, Helsinki60, Karolinska, Radboud \cite{panda}, PESO \cite{peso_dataset} and Gleason2019 \cite{gleason2019} datasets. To demonstrate that \SA{} has not just been over-fitted to the Helsinki30 dataset used in Section \ref{sec:res_robustness}, we also train and evaluate neural networks on the Helsinki60, Karolinska and Radboud datasets. For each experiment, five neural networks are trained with either \SA{}, \RA{}, \RAHE{} or \TA{} while keeping all other hyper-parameters fixed. 

The discrimination performances of neural networks trained with \SA{}, \RA{}, \RAHE{} or \TA{} are presented in Figure \ref{fig:external_val}. Compared to the current state-of-the-art methods, neural networks trained with \SA{} achieve better or similar results on every external validation dataset. Networks trained with \RA{}, \RAHE{} or \TA{} fail to generalise to several of the external datasets, despite the fact that the same networks achieve good results on other external validation datasets. For example, networks trained on the Helsinki30 dataset (first columns) with \RA{} or \RAHE{} perform comparably to \SA{} on the Helsinki60, Karolinska, Radboud and PESO datasets, but then achieve unacceptably low discrimination performances of 0.6 to 0.7 on the Gleason2019 dataset. Here, shifted evaluation allowed us to easily detect fragility to distribution shifts, which otherwise would have been missed if Gleason2019 dataset was not available.

As expected, neural networks trained on the datasets with more label noise achieve noticeably worse performance with all augmentation methods, including \SA{}. Nevertheless, networks trained with \SA{} still maintain an acceptable performance on all evaluation datasets, whereas the other methods result in no better than random discrimination performance. This is most apparent with the Karolinska (third columns) and Radboud (last columns) training datasets, where \SA{} improves the mean discrimination performance over \RA{} and \TA{} by 0.116 to 0.196 on the Helsinki30 dataset, and by 0.099 and 0.181 on the Helsinki60 dataset.

In addition to improved performance, there is a significantly lower variance in the discrimination performances across the five trained networks when using \SA{}. When training neural networks with \SA{}, \RA{}, \RAHE{} and \TA{}, the maximum difference between the best and worst-performing networks is 0.032, 0.132, 0.086 and 0.121, respectively.

\subsection{Clinical validation}
\begin{figure*}[ht]
    \centering
    \includegraphics{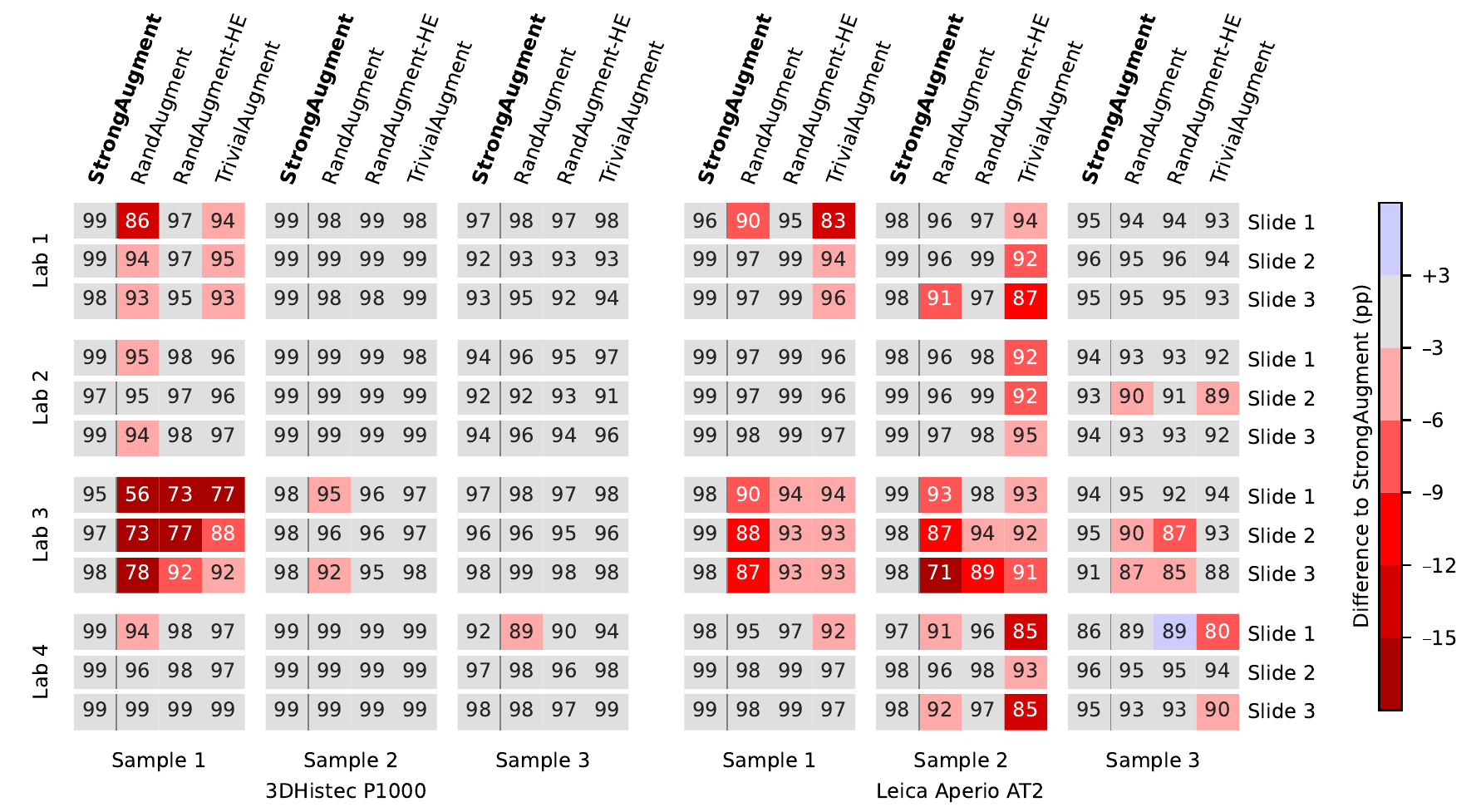}
    \caption{Discrimination performances of neural networks trained with \SA{}, \RA{}, \RAHE{} or \TA{} on 72 clinical validation cases. Each case is represented by a $1 \times 4$ block of cells, and due to different amounts of label noise, comparisons are valid only within each block. Each cell is annotated by AUROC $\times$ 100 value and coloured based on the percentage point (pp) difference to \SA{}. Neural networks trained with \SA{} perform consistently on all clinical validation cases, whereas each of the compared methods fails to generalise to several clinical validation cases.}
    \label{fig:clinical_val}
\end{figure*}

To assess neural networks' robustness to actual data distribution shifts encountered in clinical practice, we evaluate five neural networks trained on the Helsinki30 dataset with clinical validation cases described in Section \ref{sec:clinical_val}.

The discrimination performances for each of the 72 clinical validation cases are presented in Figure \ref{fig:clinical_val}. Neural networks trained with \SA{} generalise to all clinical validation cases, whereas each of the compared methods fails to generalise to several cases. \SA{} achieves at least 0.03 better AUROC values in 24, 10, and 28 cases when compared to \RA, \RAHE{} and \TA{}, respectively. \SA{} achieves at least 0.03 worse AUROC value in only a single case. Without \SA{}, neural networks fail to generalise to actual data distribution shifts encountered in clinical practice.

\subsection{Effect of spectral decoupling}

As spectral decoupling has been shown to increase robustness to data distribution shifts \cite{spectral_decoupling}, we evaluate the effect spectral decoupling has on neural network performance. Five neural networks are trained on the Helsinki30 dataset, using either \SA{}, \RA{} or \TA{}, and weight decay instead of spectral decoupling. The neural networks are then evaluated on the distribution-shifted datasets, adjusting haematoxylin and eosin stain intensities, described in Section \ref{sec:data_stain}.

The discrimination performances of the trained neural networks are presented in Supplementary Figure \ref{fig:robustness_stain_wd}. Networks trained with \SA{} achieve consistent results even without spectral decoupling, although there is a slight decrease in performance of up to four percentage points. Spectral decoupling is much more crucial for \RA{} and \TA{}, where the use of weight decay decreases the AUROC values by up to 25 percentage points. Based on these results, spectral decoupling is complementary with \SA{} and crucial for training robust neural networks.

\subsection{Other tissues}

To assess whether the previous results are somehow specific to prostate tissue, five neural networks are trained with \SA{}, \RA{} or \TA{} on the HelsinkiRCC dataset training split containing kidney tissue. The neural networks are then evaluated on distribution-shifted datasets created from the HelsinkiRCC validation split by adjusting haematoxylin and eosin stains similarly to Section \ref{sec:data_stain}.

The discrimination performances of the trained neural networks are presented in Supplementary Figure \ref{fig:robustness_stain_rcc}. Neural networks trained with \SA{} achieve consistent performance on the distribution-shifted datasets, whereas the performance of networks trained with \RA{} and \TA{} degrades quickly with even small distribution shifts. These results indicate that neural network fragility is not specific to prostate tissue, and strong augmentation is helpful with other types of tissues as well. Additionally, it is demonstrated here that shifted evaluation can also be useful when extending the validation split, and not only external datasets.

\section{Discussion}

Robustness evaluation is crucial for understanding the clinical usefulness, safety and accuracy of machine learning models intended for clinical practice \cite{to_clinic}. Despite showing strong performance on multiple external validation datasets, state-of-the-art neural networks fail to generalise to multiple clinical validation cases. With shifted evaluation, this fragility can be discovered by simply extending the validation split with distribution-shifted datasets, and addressed with strong augmentation.

Although external validations will be a crucial part of any neural network's path to clinic \cite{to_clinic}, there are serious limitations \cite{reporting_standard}. As the data-generating process is unknown, there is no way of knowing why the evaluated network fails or succeeds in external evaluation. By extending the external dataset with distribution-shifted datasets, it is possible to have known data generation functions, which completely describe any performance differences between the datasets. Thus, it is possible to discover cases where the network fails, and demonstrate robustness for the selected distribution shifts.

We believe that due to the lack of adequate evaluation methods, the role of augmentation as a method of increasing neural networks' robustness to data distribution shifts has been largely unexplored. In medical imaging, some work has been done with training stain-invariant neural networks through augmentation \cite{stain_augment, randaug_he}. A comparison of different stain augmentation and normalisation methods concluded that augmentation is crucial for achieving good performance, but found slightly better results with light rather than strong augmentation \cite{stain_norm_and_augment}. As the results were based on external validation performance and augmentation was coupled with stain normalisation, it is likely that the fragility of neural networks was simply not discovered. With natural images, the development of augmentation methods has mostly focused on improving performance on the validation split of the ImageNet \cite{imagenet} dataset \cite{autoaugment, randaug, trivaug, uniformaug}. Thus, light augmentation has been preferred over strong, which can reduce performance on in-distribution datasets, although the reduction in performance is often dwarfed when compared to the performance decreases on external datasets \cite{dl_failures}.

Interestingly, even though the augmentation space of \RA{}, \RAHE{} or \TA{} contains a given transformation, this does not guarantee that the trained network becomes robust for the distribution shift caused by the transformation. For example, the \textit{solarize} transformation is included in the augmentation spaces of \RA{} and \TA{}, but networks trained with either method are still extremely fragile for this distribution shift. \TA{} even has a wider augmentation space for the \textit{brightness} transformation, but \SA{} achieves 0.461 better AUROC when doubling the brightness of the images in the dataset. We speculate that the combination of a wider augmentation space, coupled with a significantly higher number of transformations per image makes neural networks trained with \SA{} robust to distribution shifts included in its augmentation space. Additionally, the results of this paper suggest that networks trained with \SA{} are also robust to other data distribution shifts, not just the ones included in the augmentation space.

Neural networks trained with \SA{} achieve consistent results with distribution-shifted datasets, external validation and clinical validation cases. This is not the case for current state-of-the-art methods, which are extremely fragile to distribution shifts and fail to generalise to several external datasets and clinical validation cases. The performance decreases for networks trained with \RA{}, \RAHE{} and \TA{} are unacceptable with most of the datasets, especially when considering that the decrease does not occur with networks trained with \SA{}. The magnitude of differences we show between \SA{} and current state-of-the-art methods are rarely reported. For example, \TA{} is reported to increase AUROC by 0.005 over \RA{}, and \RAHE{} \cite{randaug_he} to increase the AUROC by 0.006 over the compared method \cite{stain_norm_and_augment}. In comparison, \SA{} increases AUROC by at least 0.1 in 104, 29 and 102 datasets for \RA{}, \RAHE{} and \TA{}, respectively.

Although neural networks trained with \SA{} achieve unprecedented results, it does not solve the generalisation problem in medical imaging on its own. Other methods which increase neural networks' robustness to data distribution shifts should be used. For example, spectral decoupling has been shown to increase robustness to data distribution shifts \cite{spectral_decoupling}, and it is highly complementary with \SA{} (Figure \ref{fig:robustness_stain_wd}). Also, high-quality and representative datasets are crucial for training neural networks intended for clinical practice. This is demonstrated in Figure \ref{fig:external_val}, where datasets with more label noise achieve lower discrimination performance on external validation. With \SA{} this gap is significantly smaller, emphasising the usefulness of strong augmentation on lower-quality datasets. With \SA{}, the collection of representative training datasets can now focus on biological representativeness, instead of using multiple different scanners or staining pipelines.

\section{Conclusions}

In this study, we have exposed the fragility of neural networks using shifted evaluation. Shifted evaluation can be used to thoroughly evaluate a neural network while providing a measure of robustness to distribution shifts, and guidance on how to improve the network in case of failure. To address the exposed fragility of neural networks, we developed an augmentation strategy which allows neural networks to be robust to data distribution shifts from the training data. \SA{} achieves consistent results on external datasets and clinical validation cases, even when the current state-of-the-art methods achieve no better than random performance. The results of this study indicate that without strong augmentation, neural networks are not likely to generalise to variation encountered in clinical practice.

Our study provides a baseline for evaluating and training neural networks intended for clinical practice. Strong augmentation allows networks to perform consistently despite real-world variability, and shifted evaluation can be used to demonstrate this consistency. Although we establish the effectiveness of these methods in medical imaging, both methods are applicable to any domain.

\section*{Acknowledgements}
This work was supported by Cancer Foundation Finland [304667, 191118], Jane and Aatos Erkko Foundation [290520], Academy of Finland [322675] and Hospital District of Helsinki and Uusimaa [TYH2018214, TYH2018222, TYH2019235, TYH2019249]. The authors also wish to acknowledge Harry Nisen and Petrus Järvinen for Helsinki RCC database management, FIMM Digital Microscopy and Molecular Pathology Unit supported by HiLIFE and Biocenter Finland for imaging services, and CSC – IT Center for Science, Finland for generous computational resources.

\section*{Ethics statement}
Ethical approvals for the use of human tissue material and clinicopathologic data were obtained from Institutional Ethics Committee of Hospital District of Helsinki and Uusimaa (§70/16.5.2018; HUS/419/2018) and by the National Supervisory Authority for Welfare and Health (VALVIRA, D:no V/38176/2018). According to the national and European Union legislation on noninterventional medical research, the study was conducted without informed individual patient consents by permission of the Hospital District of Helsinki and Uusimaa (§105/21.12.2018; HUS/419/2018). The experiments conformed to the principles set out in the WMA Declaration of Helsinki and the Department of Health and Human Services Belmont Report.

\section*{Declaration of competing interests}
The authors have no interests to declare.


\section*{Credit authorship contribution statement}
\textbf{Joona Pohjonen:} Conceptualisation; Formal analysis; Visualisation; Software; Methodology; Data curation; Roles/Writing - original draft \textbf{Carolin Stürenberg:} Data curation, Writing - review \& editing; \textbf{Atte Föhr:} Data curation, Formal analysis; \textbf{Reija Randen-Brady:} Data curation; \textbf{Lassi Luomala:} Data curation; \textbf{Jouni Lohi:} Data curation; \textbf{Esa Pitkänen:} Funding acquisition; Resources; Supervision; Writing - review \& editing; \textbf{Antti Rannikko:} Funding acquisition; Resources; Supervision; Writing - review \& editing; \textbf{Tuomas Mirtti:} Conceptualisation; Data curation; Funding acquisition; Resources; Supervision; Writing - review \& editing;

\bibliographystyle{unsrt}  
\bibliography{main}

\setcounter{table}{0}
\setcounter{figure}{0}
\renewcommand{\thetable}{S\arabic{table}}
\renewcommand{\thefigure}{S\arabic{figure}}
\clearpage
\onecolumn
\section*{Supplementary Figures}
\vfill
\begin{figure}[h]
    \centering
    \includegraphics{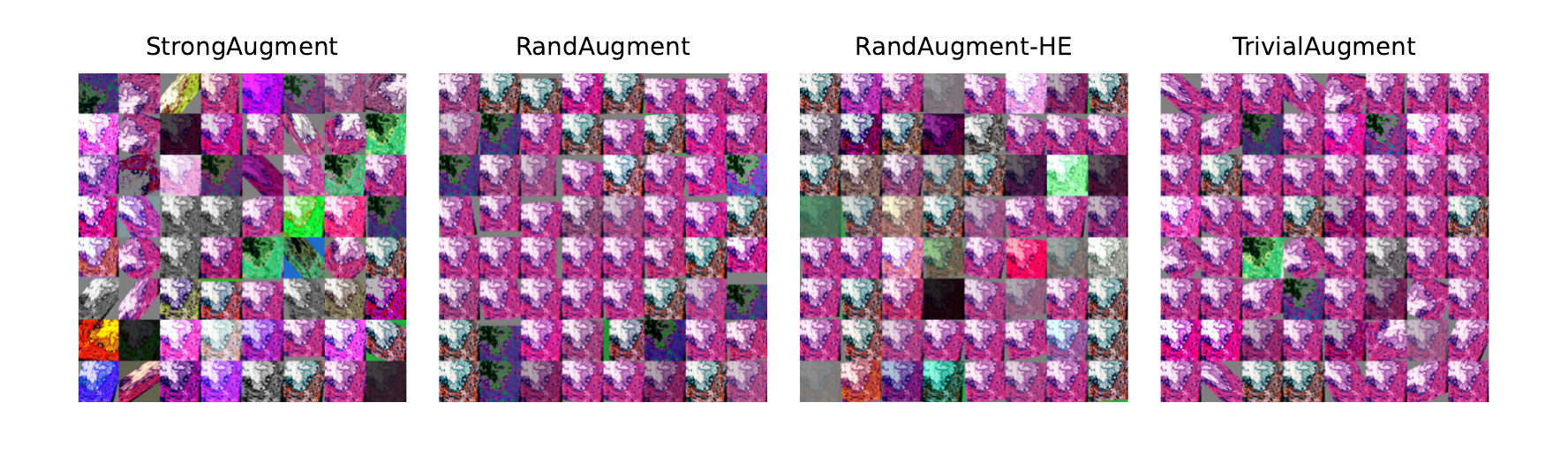}
    \caption{Examples of augmented images using \SA{} with $p=0.4$, \RA{} with $m=10$ and $n=2$, \RAHE{} with $m=5$ and $n=3$ and \TA{}. Images augmented with \SA{} contain significantly more variability than any of the other methods.}
    \label{fig:augment_comparison}
\end{figure}
\vfill
\clearpage
\begin{figure}
    \centering
    \includegraphics{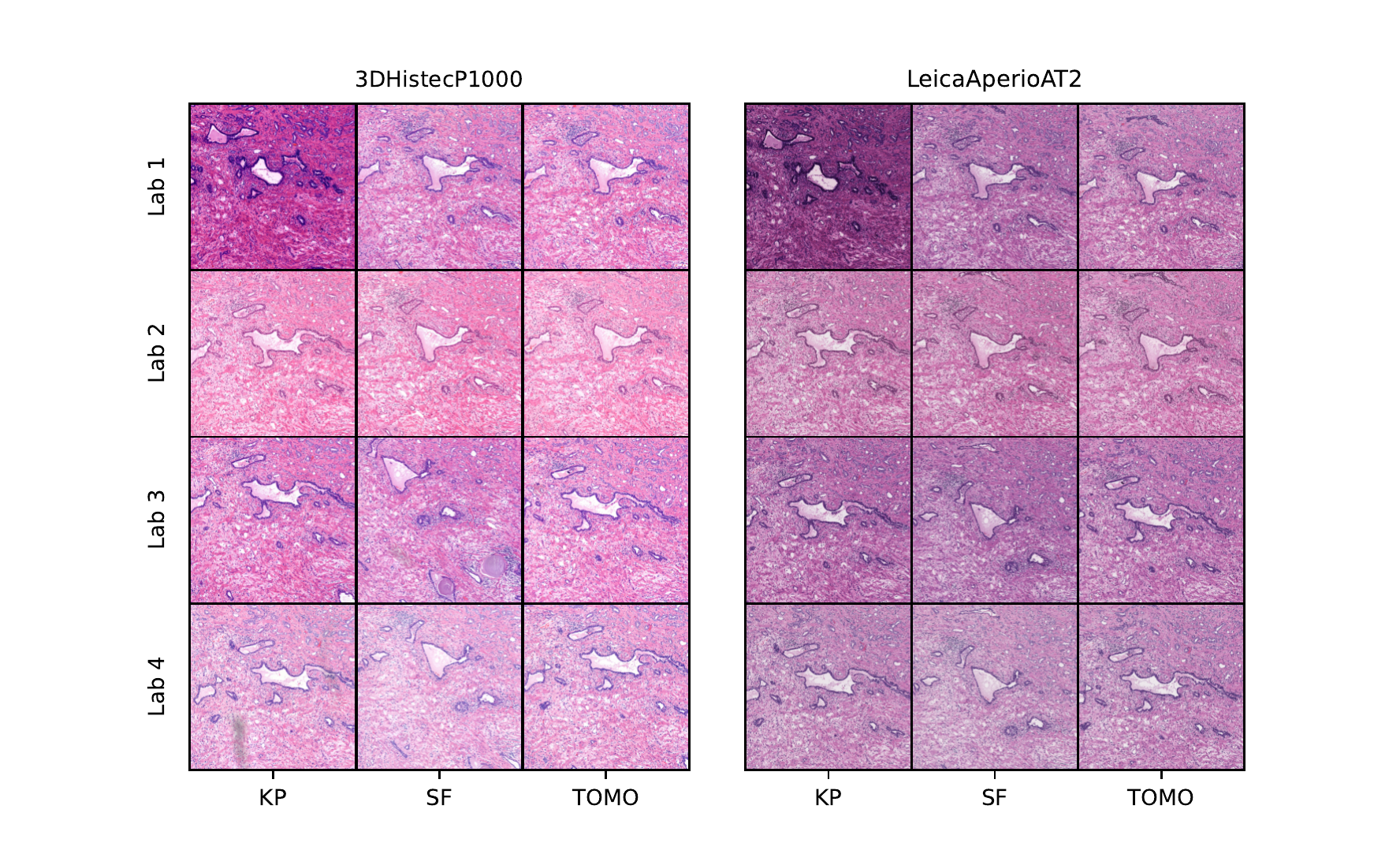}
    \caption{Same region of interest from sample 1 serial sections. There is considerable variation in the scanned images due to variability between medical centres, slide glasses and scanners. Nevertheless, neural networks intended for clinical practice should achieve consistent performance on each clinical validation case.}
    \label{fig:clinval_slides}
\end{figure}
\clearpage
\begin{figure}
    \centering
    \includegraphics{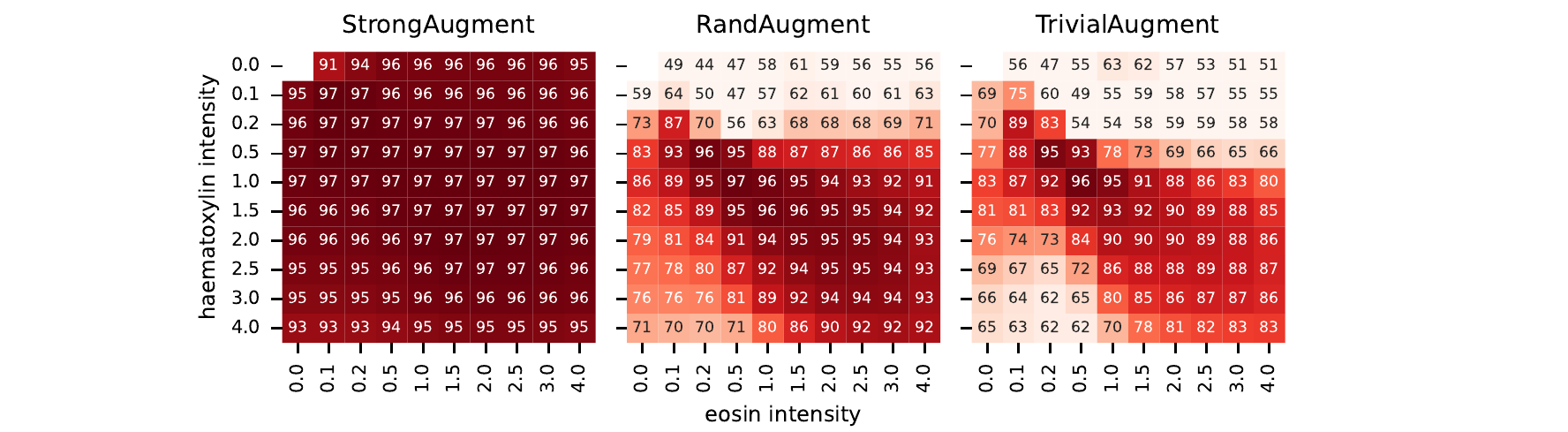}
    \caption{Robustness of neural networks, trained on the Helsinki30 dataset with weight decay instead of spectral decoupling, to distribution shifts caused by differences in the haematoxylin and eosin stain intensities. Each cell represents a given distribution-shifted dataset $\textit{stain}(X_{\text{PESO}}, h, e)$, where $h$ and $e$ denote the magnitudes for haematoxylin and eosin stains, respectively. Cells are annotated with mean AUROC $ \times$ 100 value for the five trained networks, rounded to the closest integer. Even without spectral decoupling, \SA{} achieves consistent results for each dataset. Networks trained with \RA{} or \TA{} are more dependent on spectral decoupling and achieve up to 25 percentage points lower AUROC than in Figure \ref{fig:robustness_stain}.}
    \label{fig:robustness_stain_wd}
\end{figure}
\begin{figure*}
    \centering
    \includegraphics{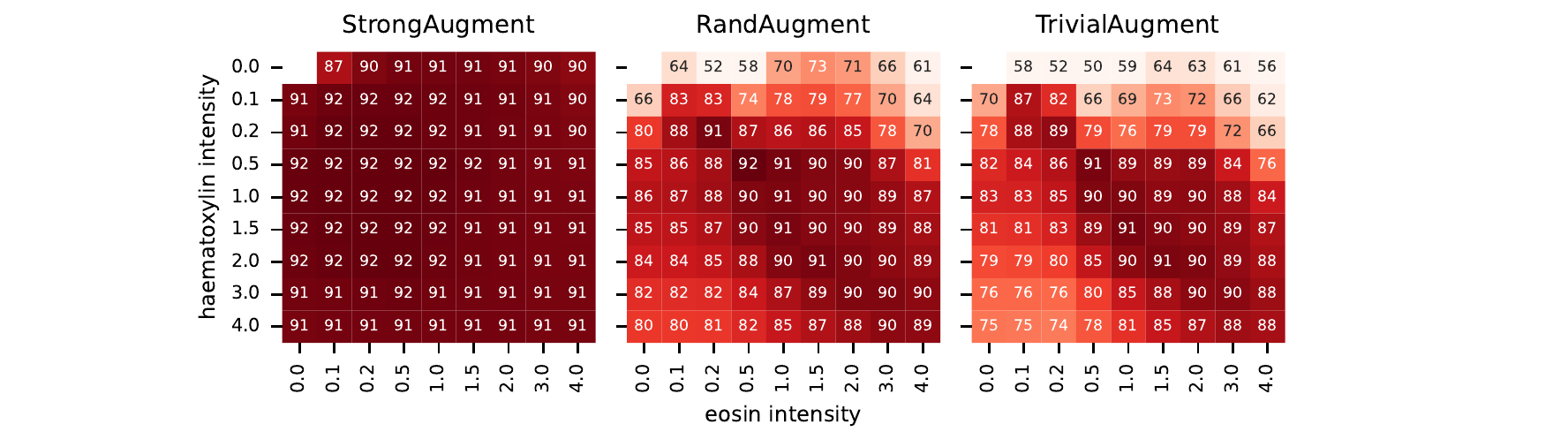}
    \caption{Robustness of neural networks, trained on the HelsinkiRCC dataset with either \SA{}, \RA{}, or \TA{}, to distribution shifts caused by differences in the haematoxylin and eosin stain intensities. Each cell represents a given distribution-shifted dataset $\textit{stain}(X_{\text{val}}, h, e)$, where $h$ and $e$ denote the magnitudes for haematoxylin and eosin stains, respectively, and $x_{\text{val}}$ the validation split of HelsinkiRCC dataset. Cells are annotated with mean AUROC $ \times$ 100 value for the five trained networks, rounded to the closest integer. Neural networks trained with \SA{} retain performance on all datasets, even when networks trained with \RA{} or \TA{} quickly degrade to no better than random discrimination.}
    \label{fig:robustness_stain_rcc}
\end{figure*}

\end{document}